\documentclass{PoS}

\title{H.E.S.S. observations of PSR B1259-63 during its 2014 periastron passage }

\ShortTitle{H.E.S.S. observations of PSR B1259-63 during its 2014 periastron passage }

\author{\speaker{C. Romoli} $^a$, P. Bordas$^b$, C. Mariaud$^c$, T. Murach$^d$, F. Aharonian$^a$ $^b$, M. de Naurois$^c$, G. P\"{u}hlhofer$^{e}$, U. Schwanke$^d$, B. van Soelen$^f$, I. Sushch$^{g}$ and V. Zabalza$^{h}$ for the H.E.S.S. Collaboration\\
\llap{$^a$} Dublin Institute for Advanced Studies, 31 Fitzwilliam Place, Dublin 2, Ireland \\
\llap{$^b$} Max-Planck-Institut f\"ur Kernphysik, P.O. Box 103980, D 69029 Heidelberg, Germany \\
\llap{$^c$} Laboratoire Leprince-Ringuet, Ecole Polytechnique, CNRS/IN2P3, F-91128 Palaiseau, France \\
\llap{$^d$} Institut f\"{u}r Physik, Humboldt-Universit\"{a}t zu Berlin, Newtonstr. 15, D 12489 Berlin, Germany\\
\llap{$^e$} Institut f\"ur Astronomie und Astrophysik, Universit\"at T\"ubingen, Sand 1, D 72076 T\"ubingen, Germany \\
\llap{$^f$} Department of Physics, University of the Free State,  PO Box 339, Bloemfontein 9300, South Africa \\
\llap{$^g$} Centre for Space Research, North-West University, Potchefstroom 2520, South Africa \\
\llap{$^h$} Department of Physics and Astronomy, The University of Leicester, University Road, Leicester, LE1 7RH, United Kingdom \\
E-mail: \email{romolic@cp.dias.ie}}
        
\abstract{An extended observation campaign of the gamma-ray binary system PSR B1259-63 has been conducted with the H.E.S.S. (High Energy Stereoscopic System) II 5-telescope array during the system's periastron passage in 2014. We report on the outcome of this campaign, which consists of more than 85 h of data covering both pre- and post-periastron orbital phases. The lower energy threshold of the H.E.S.S. II array allows very-high-energy (VHE; $E \gtrsim 100$ GeV) gamma-ray emission from PSR B1259$-$63 to be studied for the first time down to 200 GeV. The new dataset partly overlaps with and extends in phase previous H.E.S.S. campaigns on this source in 2004, 2007 and 2011, allowing for a detailed long-term characterisation of the flux level at VHEs. In addition, the 2014 campaign reported here includes VHE observations during the exact periastron time, $t_{\rm per}$, as well as data taken simultaneously to the gamma-ray flare detected with the {\it Fermi}-LAT. Our results will be discussed in a multiwavelength context, thanks to the dense broad-band monitoring campaign conducted on the system during this last periastron passage.}

\FullConference{The 34th International Cosmic Ray Conference \\
		 30 July- 6 August, 2015\\
		 The Hague, The Netherlands}

\begin{document}

\section{Introduction}

PSR B1259-63 is a gamma-ray binary system composed of a pulsar in orbit around the Be star LS 2883. PSR B1259-63 represents the only case in this specific class of systems for which the nature of the compact object, a neutron star with a spin period of 48 ms and a spin-down luminosity of $\dot{E} \approx 8 \times 10^{35}$ erg/s, has been established [1,2]. The star LS~2883 has a mass of $\sim$30 M$_{\odot}$ and drives stellar winds and an equatorial outflow that form a disc around the star \cite{1992ApJ...387L..37J,2011ApJ...732L..11N}. The orbital period of PSR B1259-63 is rather large compared with other gamma-ray binaries, about 3.4 years, and in the course of its orbit the relative distance from the Be star to the pulsar ranges from 13.4 astronomical units (AU; 1AU$\sim 1.5 \times 10^{13}$~cm) at apastron to less than 1 AU at periastron \cite{1992ApJ...387L..37J,1998MNRAS.298..997W,2004MNRAS.351..599W}. When the pulsar is close to the periastron, its interaction with the stellar and wind and  circumstellar disk gives rise to an enhancement of the non-thermal emission that is visible across the full electromagnetic spectrum, from radio to gamma rays (see, e.g., Fig.~1 in \cite{2014MNRAS.439..432C}). The high eccentricity of the orbit ($e \sim 0.87$) and an inclination angle between 19 and 31 degrees \cite{2013A&ARv..21...64D}, allow the pulsar to cross twice the circumstellar disc when approaching to and leaving from the exact time of periastron $t_{\rm per}$. In X-rays a double-peaked structure is evident in the light curve in correspondence of the two disc crossing \cite{2014MNRAS.439..432C}.

At very high energies (VHE; $E \gtrsim 100$ GeV) the High Energy Stereoscopic System (H.E.S.S.) collaboration discovered the source during the periastron passage in 2004 and recorded also the subsequent passages in 2007 and 2010/2011, whereas observations far from the periastron passage returned only flux upper limits \cite{2005A&A...442....1A,2009A&A...507..389A,2013A&A...551A..94H}. Due to visibility constraints, for each epoch a different period of the periastron passage was covered in these campaigns. A combined phase-folded analysis of these data nevertheless showed a hint of a double-peak profile in the TeV data similar to that observed in X-rays, with a similar minimum at $\sim t_{\rm per}$ \cite{2013A&A...551A..94H}. 

In the high energy (HE; E > 100 MeV) range the source was be detected by the Large Area Telescope (LAT) onboard of the \emph{Fermi} satellite during the 2010/2011 passage. On that occasion a high-energy flare was observed about 30 days after $t_{\rm per}$, releasing almost all of the spin-down luminosity in gamma rays. This flare appeared as a surprise due to the lack of any correlated flux enhancement observed at other wavelengths \cite{2011ApJ...736L..11A}.

In this manuscript we report the results of the H.E.S.S. observation campaign on the source during the last periastron passage in 2014. In this case the visibility from the H.E.S.S. site was particularly favourable for the observation of the source during the first disc crossing, the actual periastron passage and the second disc crossing, with also a good coverage of the expected HE gamma-ray flare. In section \ref{sec:obs} we shortly describe the H.E.S.S. experiment in its phase II and the observation strategy adopted. In section \ref{sec:analysis} we give a brief overview of the data analysis and in section \ref{sec:results} we show the results of the campaign comparing it with the previous recorded events. We are also reporting the the results of the analysis of {\it Fermi}-LAT as well as X-ray satellite observations of the system in 2014 taken contemporaneously to the H.E.S.S. observations.

\section{H.E.S.S. observations in 2014}
\label{sec:obs}

H.E.S.S. in its initial phase consisted of 4 Cherenkov telescopes with a diameter of 12 meters installed in the Khomas Highland in Namibia at 1800 m above the sea level. In 2012 the array entered in the phase II with the installation of a fifth telescope (CT5) with 28 meters diameter allowing the telescope to reach an energy threshold as low as $\sim$20 GeV for pulsar analysis \cite{vela_proc}. The observation of the source made use of the full array. In this work we are presenting results with a {\it conservative} energy threshold of 200 GeV to avoid possible systematic issues and background contamination effects that can affect the robustness of the analysis. This is almost a factor 2 lower than the threshold reached by the phase I of the experiment in 2004 \cite{2005A&A...442....1A}. We note that this value is not yet optimized for the new phase of the experiment and further investigation at lower energies (E<200 GeV) is being performed and will be reported in a follow-up paper.

The H.E.S.S. 2014 observation campaign on PSR B1259-63 started as early as of March 2014 and lasted until late July 2014. The source was monitored almost daily, with observations prevented only during bright/full-moon nights and when bad weather or technical problems didn't allow a safe telescope operation. At the H.E.S.S. site, the source culminates at relatively low altitudes, and the average zenith angle for the whole campaign was $\sim 42 \deg$.

%The visibility of the source for the 2014 passage is shown in Fig.~\ref{fig:visibility}, in comparison with the previous H.E.S.S. observations and with the flare detected in 2011 by the \emph{Fermi}-LAT. 

%\begin{figure}
%\centering
%\includegraphics[scale=0.25]{visibility-plot-psrb.png}
%\caption{Visibility from the H.E.S.S. site of PSR~B1259-63 during the 2014 observation campaign compared with the rescaled light-curves of the previous passages and of the flare detected by the \emph{Fermi}-LAT. The HESS data points are from \cite{2013A&A...551A..94H} while the \emph{Fermi}-LAT lightcurve was taken from \cite{2011ApJ...736L..11A}.}
%\label{fig:visibility}
%\end{figure}

%The observations with the Cherenkov array were accompanied by a multiwavelength coverage with data in the X-ray and optical range.
%SHOULD WE PROVIDE ALSO THE X-RAY AND OPTICAL LIGHTCURVE???

\section{Data Analysis}
\label{sec:analysis}

\subsection{H.E.S.S. Analysis}

\begin{figure}
\centering
\includegraphics[scale=0.25]{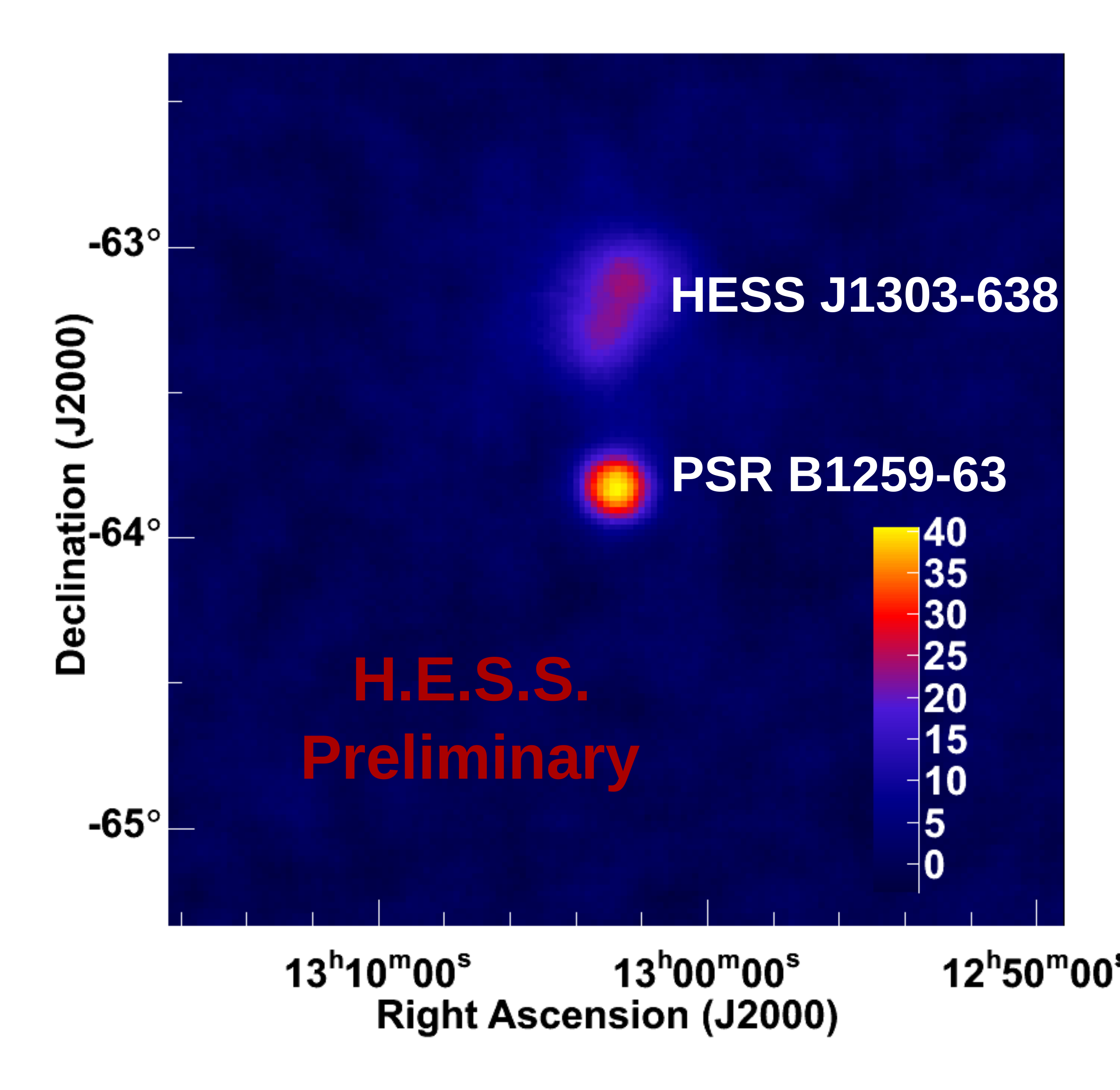}
\caption{Significance map of the sky region around PSR~B1259$-$63 obtained with the \emph{stereo} analysis reconstruction chain (see text for details). The source above PSR~B1259$-$63 is HESS J1303$-$631, a pulsar wind nebula associated with the pulsar PSR~J1303$-$6305 \cite{2012A&A...548A..46H}.}
\label{fig:sign_map}
\end{figure}

The H.E.S.S. data analysis has been performed with the {\it Model} reconstruction method \cite{2009APh....32..231D} adapted to work also with the new telescope CT5 data. All the results have been cross-checked with at least one independent reconstruction chain either with analysis chains making use of neural networks reconstruction techniques based on initial set of Hillas parameters \cite{thomas_proc}, or with and adapted version of the ImPACT (Image Pixel-wise fit for Atmospheric Cherenkov Telescopes; \cite{2014APh....56...26P}) method. In order to explore the low-energy range, data were analysed in the so-called {\it mono} configuration, for which events triggered by the CT5 telescope alone are considered. To study the high-energy range and allow instead for a comparison of the VHE flux obtained in 2014 with that reported in previous periastorn passages, we perfomed a {\it stereo} reconstruction, using simultaneous events involving CT5 and at least one of the other 4 telescopes. Once run selection and data quality cuts are applied, the dataset consists of 57.4 hours of livetime. An image displaying the significance distribution around PSR B1259$-$63 is shown in Fig.~\ref{fig:sign_map}.

\subsection{{\it Fermi}-LAT Analysis}

The {\it Fermi}-LAT data were analysed in the time period between the 1st of March and the 30th of July 2014, so to cover the full periastron passage. The analysis was performed with the \emph{Fermi Science Tools} provided by the {\it Fermi} collaboration\footnote{{\tt http://fermi.gsfc.nasa.gov/ssc/data/analysis/}} using the version of the software v9r32p5 with "P7REP\_SOURCE\_V15" IRFs. The data were treated according to the {\it Fermi}-LAT guidelines and a binned likelihood fit was done on a region of interest of 30$^{\circ}$x30$^{\circ}$ taking into account all the sources in the 2FGL catalogue \cite{2012ApJS..199...31N} within a radius of 25 degrees from PSR~B1259$-$63 fixing their spectral parameters in case their distance from our source was greater than 5 degrees. Using the parameters derived from this fit, a weekly light curve was extracted over the full period. A daily light curve was computed in addition only for the 5 weekly bins with significant detection. The likelihood fit of this interval gave as a result an average flux above 100 MeV of $\left( 1.0 \pm 0.5 \right)\times 10^{-6}$~ph~cm$^{-2}$~s$^{-1}$ and a photon index of $3.02 \pm 0.08$. The two light curves are shown in the middle panel of Fig.~\ref{fig:mono_mwl}.

\subsection{X-ray analysis}

During 2014 the {\it Swift}-XRT performed 29 observations of PSR~B1259$-$63.
Cleaned event files were obtained using the \texttt{xrtpipeline} task from
HEAasoft v6.15.1. For each observation, a source spectrum was extracted from a
circular region of 1\,arcmin around the nominal position of PSR~B1259$-$63, and a
background spectrum extracted from a nearby region of the same size devoid of
sources.  Xspec v12.8.1 was used to perform a spectral fitting of the spectra
and obtain the 1--10 keV flux. An absorbed powerlaw model was used for the
fitting, and given the relatively low count number in the XRT spectra, the
hydrogen column was fixed to $N_\mathrm{H}=5\times10^{21}$\,cm$^2$ as in \cite{2014MNRAS.439..432C}. The flux light curve for these observations can be seen in the lower panel of Fig.~\ref{fig:mono_mwl} along with X-ray measurements made in previous years with \emph{Swift}-XRT and other X-ray observatories.

\section{Results}
\label{sec:results}

\begin{figure}
\centering
\includegraphics[scale=0.25]{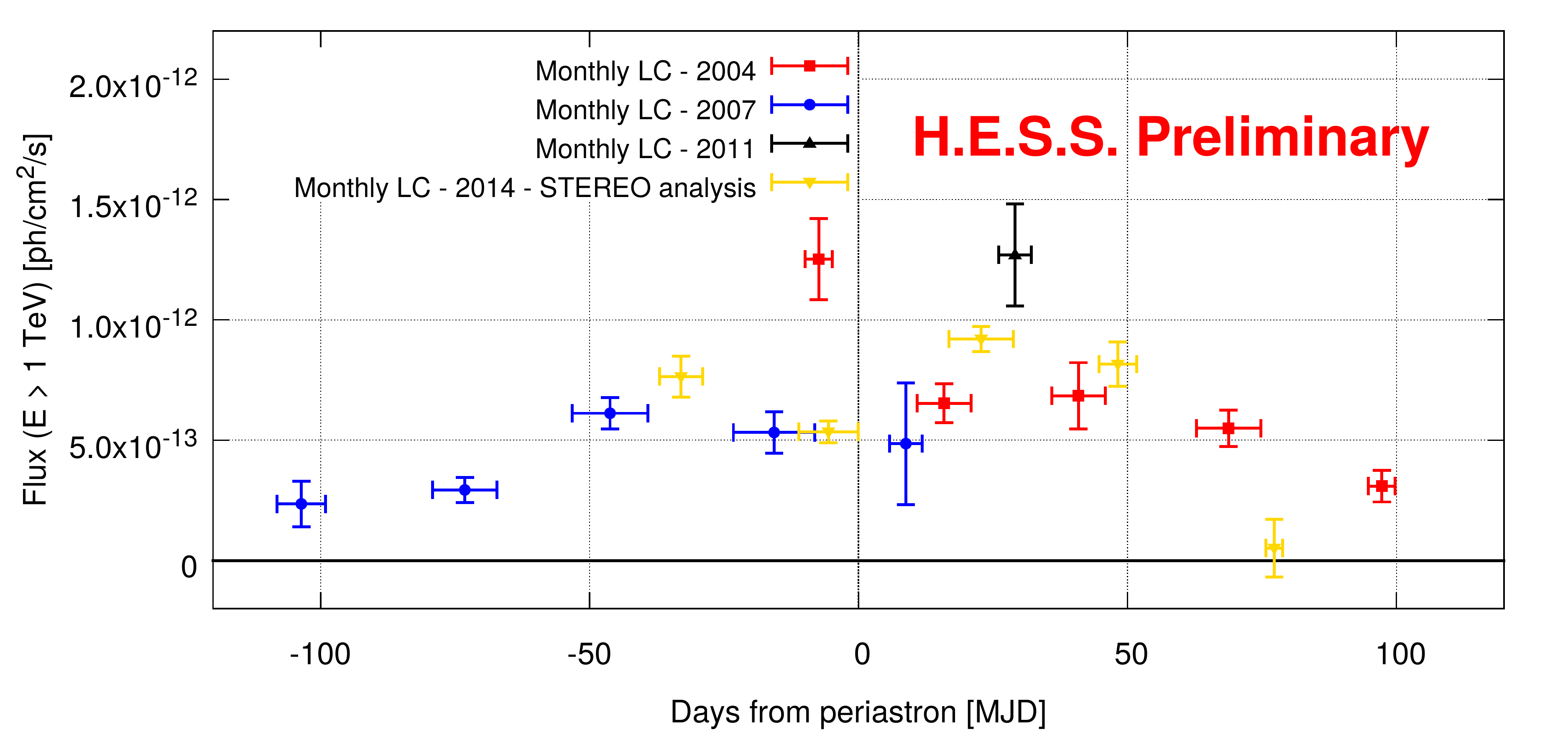}
\caption{Monthly light curve of the previous periastron passages (2004, 2007 and 2011) with the \emph{stereo} light curve of the 2014 event with TeV integral fluxes above 1 TeV.}
\label{fig:full_periodwise}
\end{figure}

\begin{figure}
\centering
\includegraphics[scale=0.25]{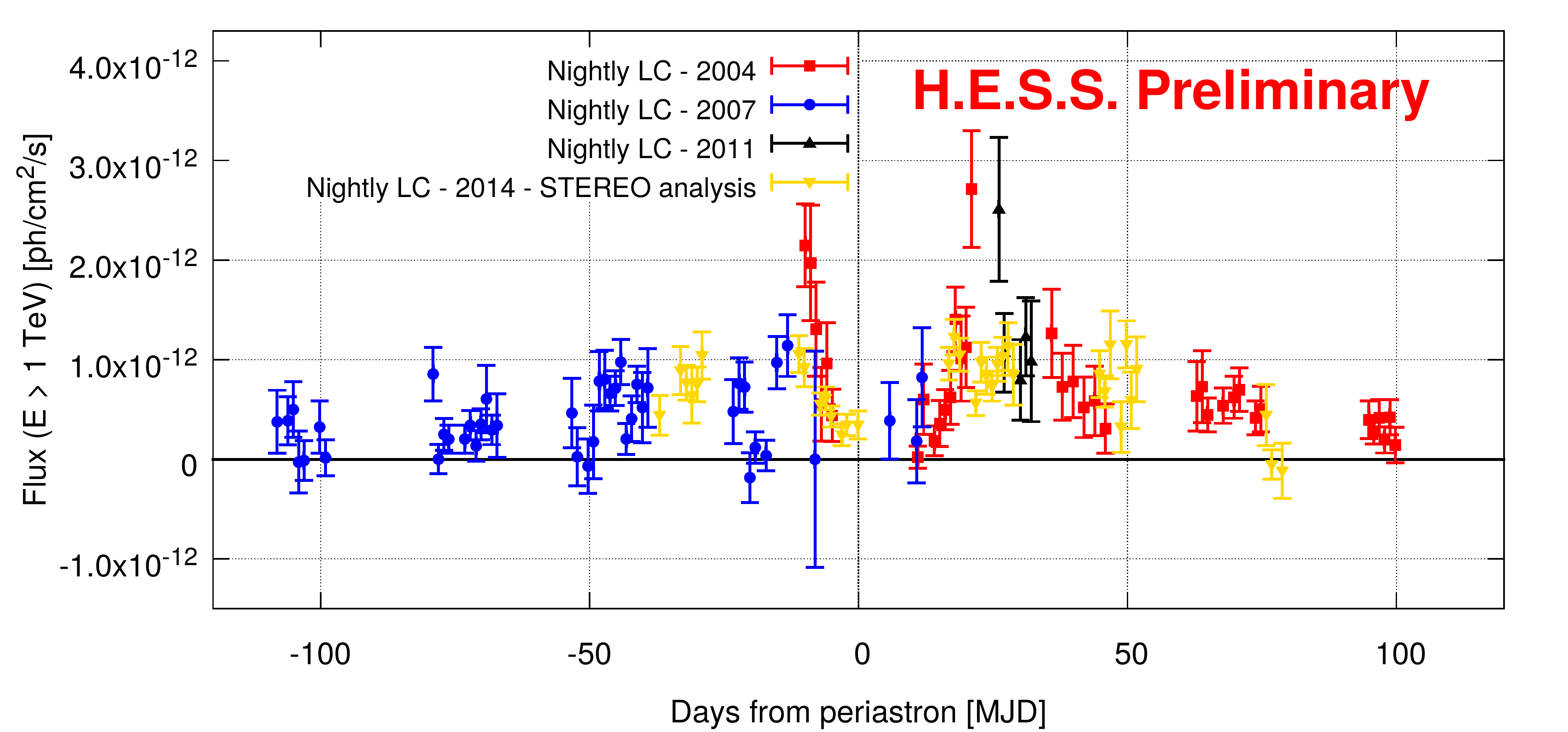}
\caption{Nightly light curve for the 2004, 2007, 2011 and 2014 datasets. As in figure {\protect\ref{fig:full_periodwise}} the fluxes are reported for photon energies above 1 TeV. The 2014 data points were obtained with the {\it stereo} analysis reconstruction technique (see text for details).}
\label{fig:full_nightwise}
\end{figure}

In Fig.~\ref{fig:full_periodwise} and Fig.~\ref{fig:full_nightwise} we are presenting the monthly and nightly light curves of the source recorded by H.E.S.S. 
together with observations obtained during the previous periastron passages in 2004, 2007 and 2010/2011. As a consistency check, we also performed a re-analysis of these previous observations with the more recent and advanced analysis tools available. This reanalysis did not yield any significant change compared to the published results in \cite{2013A&A...551A..94H}. The 2014 data were analysed using the \emph{stereo} reconstruction to compare results from similar reconstruction techniques and in the same energy range.

The monthly-folded light curve in Fig.~\ref{fig:full_periodwise} displays a significant difference in the pre-periastron period between the average flux recorded in 2004 and those obtained in 2007 and 2014, with the 2004 observations being a factor of about two higher than the 2007 and 2014 ones, with a distance\footnote{computed as the compatibility with 0 of the difference between the two values with the respective uncertainties taken into account.} between them at a statistical level of 4$\sigma$. The nightly-averaged light curve displayed in Fig.\ref{fig:full_nightwise} provides, on the other hand, evidence for variability on short time scales. In particular, the observations taken in 2014 show the declining emission of PSR B1259-63 when the system approaches the exact time of periastron $t_{\rm per}$, which represents a local minimum in the double-peaked nightly-averaged light curve profile.

\begin{figure}
\centering
\includegraphics[scale=0.22]{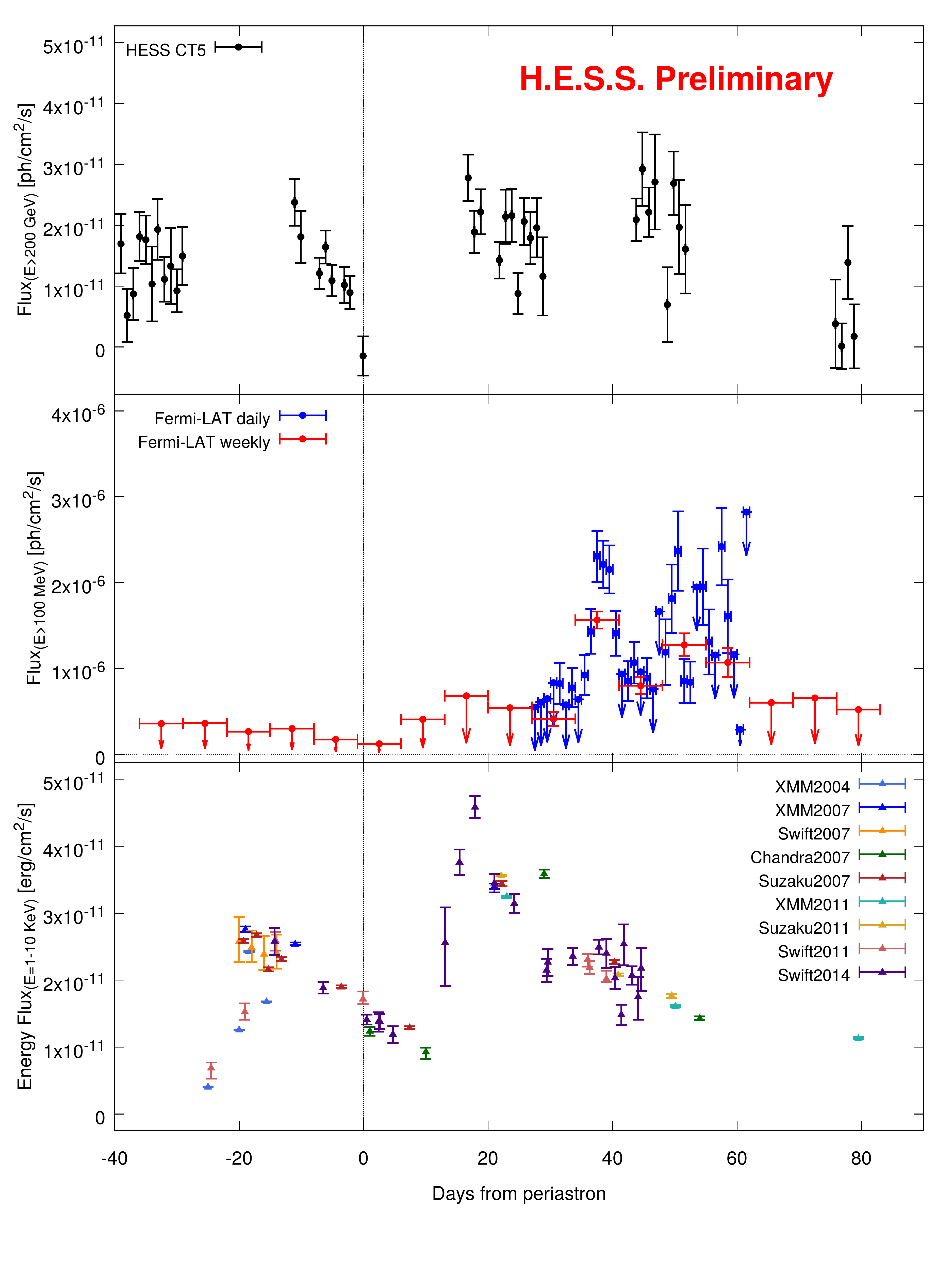}
\caption{{\bf Top panel:} CT5 \emph{mono} light curve above 200~GeV. {\bf Middle panel:} flux obtained from the analysis of {\it Fermi}-LAT data in the energy range 100~MeV - 300~GeV. The points of the {\it Fermi}-LAT light curve were computed if the TS value of the source was above 10 ($\sim3\sigma$). A $2\sigma$ upper limit was computed otherwise. {\bf Lower panel:} energy flux measured with {\it Swift}-XRT in 2014 together with archival data obtained with {\it XMM-Newton}, $Chandra$ and $Suzaku$ during previous periastron passages.}
\label{fig:mono_mwl}
\end{figure}

Due to the presence of CT5 it is also possible to explore the lower energy behaviour of the source, down to 200 GeV, and to look for the presence of a flux enhancement in correspondence of the of the high-energy flare detected again in 2014 \cite{2014ATel.6225....1W}. In Fig.~\ref{fig:mono_mwl} we report the CT5 light curve in comparison with the {\it Fermi}-LAT emission above 100 MeV and the energy flux at X-ray as measured with {\it Swift}-XRT, in the energy range 1 - 10 keV, together with the results obtained in previous periastron passages. Due to moon constraints, it was not possible to cover the onset of the HE gamma-ray flare. TeV observations during this period nevertheless  covered orbital phases in which the source was active at HE. Despite no evidence of a flaring event at VHE, the source displayed a relatively high flux state, with average values close to those observed during the first and second disc crossing. Furthermore, the X-ray light curve after 20 days from periastron shows a behaviour more complex than a simple exponential decay of the flux, with hints of variability around 40 days after the periastron passage \cite{2014ATel.6248....1B}. 

\begin{figure}
\centering
\includegraphics[scale=0.3]{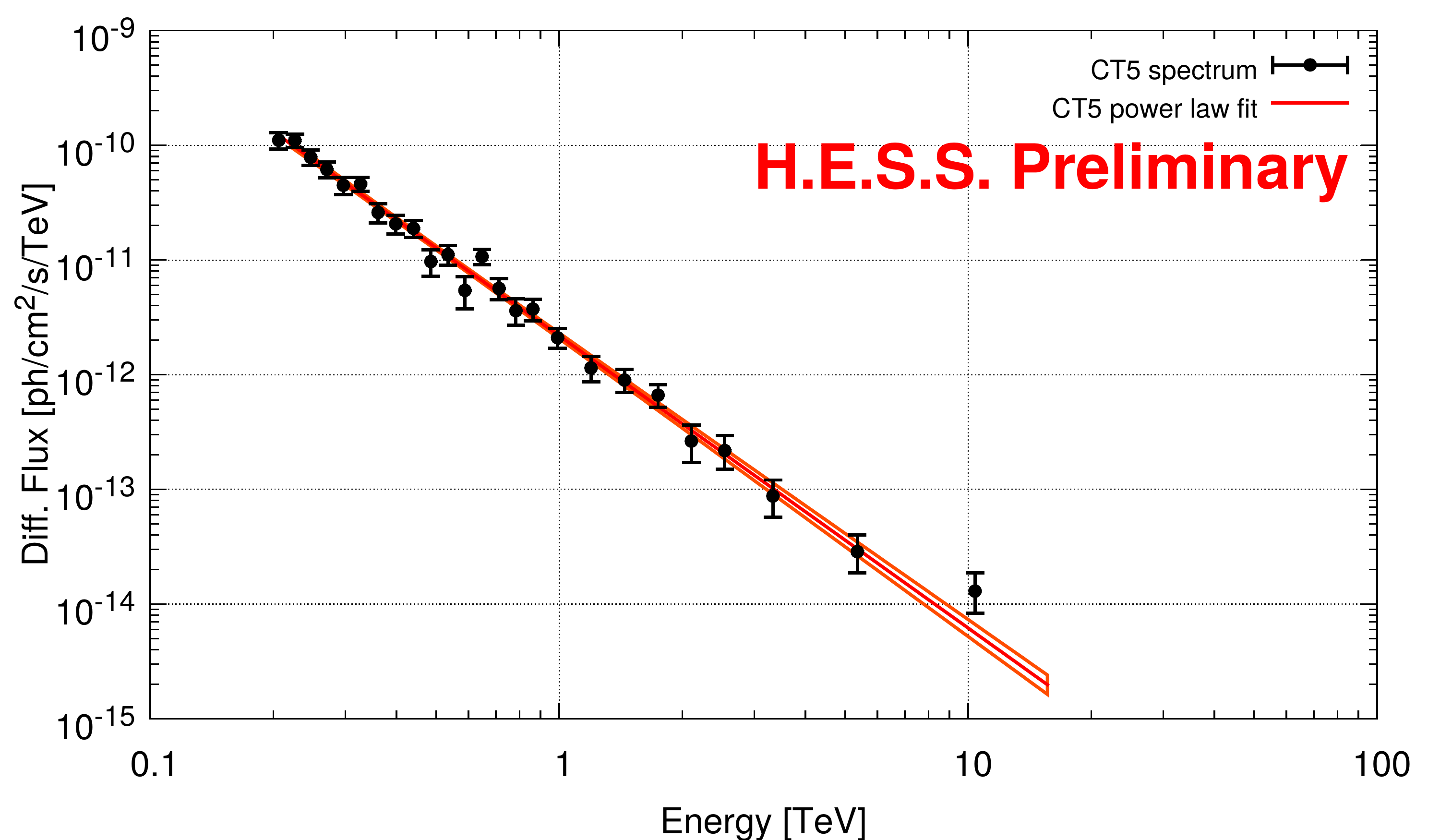}
\caption{H.E.S.S. CT5 spectrum for the 2014 data. The power law fit is now dominated by the low energy photons resulting in a spectrum harder than in previous passages with a value $\alpha =2.55\pm0.06$ for E>200 GeV.}
\label{fig:mono_spectrum}
\end{figure}

We produced the spectrum using \emph{mono} data and fit it using a power law function: $\frac{dN}{dE}=N_0\left(\frac{E}{E_{ref}}\right)^{-\alpha}$ where $N_0$ is the normalization parameter and $E_{ref}$ is the reference energy. For the spectrum shown in Fig.~\ref{fig:mono_spectrum}, using $E_{ref}=0.49$~TeV, we retrieve $N_0=(1.36 \pm 0.06)\times 10^{-11}$ ph cm$^{-2}$ s$^{-1}$ TeV$^{-1}$ and a photon index $\alpha=2.55 \pm 0.06$. This last value is harder but still within 2$\sigma$ compared to previous results. We did not find any evidence of a break feature or curvature in the energy range 200~GeV~$-$~10~TeV.

\section{Summary}

In this proceedings we have presented the preliminary results of an observation campaign conducted in 2014 on the gamma-ray binary PSR~B1259$-$63 making use for the first time of the advanced capabilities of the new H.E.S.S. phase-II array, allowing for the study of the source down to energies of 200 GeV. The good H.E.S.S. observation conditions of the source in 2014 permitted an almost complete coverage of all pre- and post-periastron orbital phases, as well as observations right at $t_{\rm per}$, which display a smooth flux decline towards a relative minimum of the TeV emission, similar to the behavior observed at lower energies. Furthermore, the good phase coverage permits a direct inspection of the long-term (super-orbital) variability of the VHE flux by comparing the phase-folded fluxes recorded in a relatively long time scale of $\sim 10$~years. In addition, part of the 2014 observations have been taken contemporaneously to a HE gamma-ray flare observed with the $Fermi$-LAT, similar to the one reported in 2011. Although the H.E.S.S. data does not provide any significant evidence of an abrupt increase of the VHE emission similar to the HE case, the new 2014 observations nevertheless reveal a relatively high source flux state at TeV energies $\sim$~50 days after $t_{\rm per}$.

Finally, we remark that the results announced in these proceedings are still preliminary. A more detailed final analysis and a full discussion of the obtained results are in preparation. 
\\
\\
{
\footnotesize
{\bf Acknowledgments}: The support of the Namibian authorities and of the University of Namibia in facilitating the construction and operation of H.E.S.S. is gratefully acknowledged, as is the support by the German Ministry for Education and Research (BMBF), the Max Planck Society, the German Research Foundation (DFG), the French Ministry for Research, the CNRS-IN2P3 and the Astroparticle Interdisciplinary Programme of the CNRS, the U.K. Science and Technology Facilities Council (STFC), the IPNP of the Charles University, the Czech Science Foundation, the Polish Ministry of Science and Higher Education, the South African Department of Science and Technology and National Research Foundation, and by the University of Namibia. We appreciate the excellent work of the technical support staff in Berlin, Durham, Hamburg, Heidelberg, Palaiseau, Paris, Saclay, and in Namibia in the construction and operation of the equipment.
}

\end{document}